\documentclass[10pt, conference]{IEEEtran}
\usepackage[noadjust]{cite}
\usepackage{comment}
\usepackage[
	colorlinks=true,
	linkcolor=blue,
	citecolor=blue,
	urlcolor=blue,
]{hyperref}
\usepackage{amsmath,amssymb,amsfonts,mathtools}
\usepackage{graphicx}
\usepackage{svg}
\usepackage{tabularx}
\usepackage{xcolor}
\usepackage[capitalize]{cleveref}
\usepackage[inline]{enumitem}
\usepackage{caption}
\usepackage{subcaption}
\usepackage{listings}
\usepackage{siunitx}
\usepackage{stfloats}
\usepackage{xparse}

\graphicspath{{./Graphics/}}

\newcommand{\pulselib}[0]{\texttt{pulselib}}
\newcommand{\Pulselib}[0]{\texttt{Pulselib}}

\begin{document}
\bstctlcite{references:BSTcontrol}
\title{
	Towards a pulse-level intermediate representation for diverse quantum control systems
}

\author{
	\IEEEauthorblockN{
		Jude~Alnas\IEEEauthorrefmark{1}\IEEEauthorrefmark{2},
		Aniket~S.~Dalvi\IEEEauthorrefmark{1} and
		Kenneth~R.~Brown\IEEEauthorrefmark{1}\IEEEauthorrefmark{3}
	}

	\IEEEauthorblockA{
		\IEEEauthorrefmark{1} Duke Quantum Center and Department of Electrical and Computer Engineering,
		Duke University, Durham, NC
	}
	\IEEEauthorblockA{
		\IEEEauthorrefmark{3} Department of Physics and Department of Chemistry,
		Duke University, Durham, NC
	}
	\IEEEauthorblockA{
		\IEEEauthorrefmark{2}Email: jude.alnas@duke.edu
	}
}

\maketitle

\begin{abstract}
	A system-independent intermediate representation (IR) for pulse-level programming of quantum control systems is required to enable rapid development and reuse of quantum software across diverse platforms.
In this work, we demonstrate the utility of \pulselib{}~\cite{Dalvi2024} as a candidate for such an IR.
We implement graph-based IRs and transpilation pipelines for two unique frequency synthesizers and benchmark performance against existing IRs.
Key elements of these pipelines are munchers and parametrizable pulse schedules.
The former encodes target-specific constraints and allows translation of fundamentally system-agnostic pulse descriptions to arbitrary low-level representations, and the latter enables schedule reuse that produces savings in transpilation time relative to device-specific alternatives.
Benchmarks reveal that \pulselib{} provides performance comparable to fast, device-specific IRs while providing a speedup of up to 4.5x over existing IRs. 
For highly parametrized applications, \pulselib{} provides favorable scaling of transpilation times with respect to the number of parameters and can exhibit speedups relative to existing IRs up to 69\% larger than speedups provided by optimized, device-specific techniques.
\end{abstract}


\section{Introduction}\label{sec:intro}

The evolution of quantum information processing from a theoretical concept to a growing industry has been driven by the innovations in robust control of quantum systems.
Precise pulse-level control in particular has led to advances in gate fidelities~\cite{Abdelhafez2020a, Kang2021}, circuit compilation~\cite{Shi2019a}, and quantum simulation~\cite{Monroe2021,Semeghini2021,Andersen2024}, where the last is especially promising for demonstrating quantum advantage~\cite{Kang2024}.
However, unique differences in qubit platforms has produced varied architectures and technology stacks for quantum control systems, leading to a fragmented ecosystem of quantum control software.
Pulse-level application code often invokes low-level interfaces that are specific to the target quantum system,
e.g. the programming interface exposed by a signal generator driving coherent qubit rotations.
The result is software that is tightly coupled to the target quantum system, making it difficult to write and reuse code across different platforms and slowing the development of new applications.

Compare this to the quantum circuit model for quantum computation, where quantum circuits provide a natural, system-independent representation of a quantum algorithm.
Quantum processors need only support a universal gate set in order to approximate to arbitrary accuracy any quantum circuit~\cite{Kitaev1997}.
Transpilers and compilers can then be used to convert a high-level description of a quantum circuit, often in domain-specific languages (DSLs) such as Qiskit~\cite{Qiskit}, into a low-level representation that is compatible with the target quantum system.

The quantum circuit model, however, is unsuitable for pulse-level applications since pulse-level details are hidden behind gate abstractions.
Therefore, a pulse-level intermediate representation (IR) is needed that provides a system-independent representation of pulse-level applications.
Additionally, such an IR should at minimum allow for the construction of complex pulse schedules in a high-level language and a pipeline for system-aware transpilation of this schedule to device-level descriptions.
A number of IRs currently exist that attempt to fulfill this role.
Qiskit Pulse~\cite{Mckay2018,Alexander2020} is a popular choice for pulse-level programming, providing mechanisms for constructing symbolic and parametrized pulse schedules and pipelines for transpiling these schedules into system- or device-specific representations.
However, Qiskit Pulse internally uses a sample-based pulse representation that is incompatible with parameter-based signal generators, and the package as a whole has recently been deprecated.
Additional choices are Pulser~\cite{Silverio2022} and JaqalPaw~\cite{Lobser2021}, but the former is explicitly tailored towards neutral atom systems, and the latter targets a single class of devices and lacks semantics for pulse scheduling.



In Ref.~\cite{Dalvi2024}, the authors proposed the system-agnostic pulse-level representation \pulselib{}, and here we demonstrate its usage as an IR and design a transpilation pipeline targeting quantum control systems with diverse control hardware.
In \cref{sec:bg}, we provide a brief overview of \pulselib{} and its design principles as well as the two signal generators used in this work: the ARTIQ AD9910~\cite{bourdeauducq_2016_51303,ad9910} and the Octet RFSoC~\cite{rfsoc,Lobser2020}.
The structure of a \pulselib{}-based transpilation pipeline translating high-level pulse descriptions to arbitrary low-level descriptions is introduced in \cref{sec:pipeline}, and we provide implementations of this pipeline for the two signal generators in \cref{sec:imp}. 
By leveraging \pulselib{}'s extensibility, we also introduce custom nodes that capture the full capabilities of the Octet RFSoC, including its support for advanced modulation and phase tracking techniques.
In \cref{sec:bench}, we compare \pulselib{}'s performance relative to existing techniques through benchmarks inspired by common applications in trapped-ion quantum systems.
These tests reveal that the overhead of using \pulselib{} is comparable or even less than that of existing IRs, particularly when using highly parametrized schedules.

The major contributions of this paper are the following:
\begin{enumerate}
    \item We identify how \pulselib{} may be used as an IR by defining a workflow in which \pulselib{} is used to both construct and transpile pulse schedules targeting diverse quantum control systems.
    \item We provide example implementations of this pipeline for two different signal generators: the ARTIQ AD9910 DDS and the Octet RFSoC.
    \item We demonstrate \pulselib{}'s extensibility by introducing custom nodes that capture the full capabilities of the RFSoC, including its support for advanced modulation and phase tracking techniques.
    \item We demonstrate the performance of our IRs through benchmarks inspired by common applications in trapped-ion quantum systems, showing that the overhead of using \pulselib{} is comparable or even less than that of existing IRs or even device-specific representations, particularly when using highly parametrized schedules.
\end{enumerate}

\section{Background}\label{sec:bg}

\subsection{Pulselib}\label{sec:bg:pulselib}
\begin{figure}
    \centering
    \includegraphics[width=.7\linewidth]{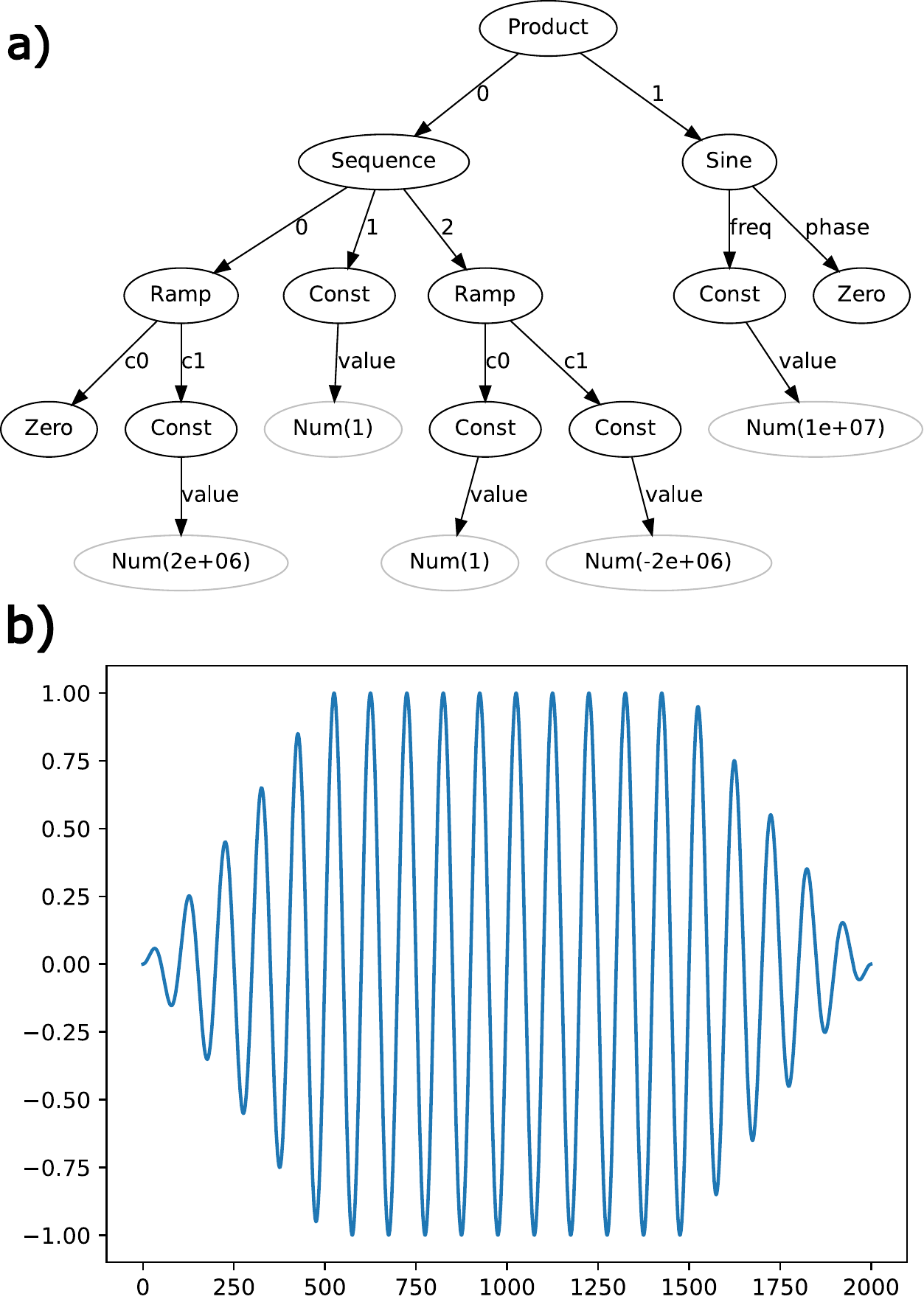}
    {\phantomsubcaption\label{fig:pl-example:graph}}
    {\phantomsubcaption\label{fig:pl-example:samples}}
    \caption{
        (a) \Pulselib{} pulse graph representing a piecewise-linear ramp of a 10 MHz tone.
        Ramp node edges labelled $c_i$ correspond to coefficients in the polynomial $y(t)=\sum_{i}c_{i} t^{i}$.
        (b) Sampled representation of the pulse graph in (a).
        \Pulselib{} provides a built-in mechanism for sampling pulse graphs, allowing for easy visualization, debugging, and usage with compatible signal generators.
    }
    \label{fig:pl-example}
\end{figure}

Pulses in \pulselib{} are internally represented as directed acyclic graphs (DAG) where nodes represent kinds of pulses (e.g., sine, Gaussian, polynomial, etc.).
Each node exposes a set of outgoing edges each of which correspond to a pulse parameter.
Edges point to the parameter value which is itself a node, allowing construction of complex pulses.
Arithmetic and sequencing operations are represented as operator nodes and can be used to combine multiple pulses into complex modulated or composite pulses, further increasing \pulselib{}'s expressibility.
\Cref{fig:pl-example} shows a simple example of a pulse graph representing a linear ramp of a 10 MHz tone.
The piecewise-linear amplitude profile is constructed as a sequence of ramps and constant segments.
This is then multiplied by a sine wave in the form of \texttt{puslselib.Sine} node, creating a \texttt{pulselib.Product} node whose children nodes are the factors.

Coupled with graph-based representations, \pulselib{} provides mechanisms for transforming pulse graphs.
Graph transformations are realized through custom implementations of the visitor design pattern~\cite{Gamma1994}.
Basic visitors perform depth-first graph traversals, calling methods on each encountered node.
These methods dispatch to the appropriate method for the node type, allowing for polymorphic behavior.
Transformers are a derivation of visitors that can alter the structure of the input graph.
These are used to make graph simplifications or to apply equivalence transformations such as the deletion of any zero duration nodes or the merging of nodes with identical parameters.

Pulse parametrization is a key feature of \pulselib{} that is enabled by the combination of the graph representation and visitor pattern.
Instead of providing concrete values as pulse parameters, \pulselib{} allows users to provide symbolic values in the form of \texttt{pulselib.Var} nodes.
Concrete values may later be assigned to these variables using a substitution visitor.
This visitor holds a mapping of variable names to values and traverses the graph, replacing encountered variables with their corresponding value.
Substitutions can also be reset, allowing for the same graph to be reused with different parameter values.

A third unique implementation of the visitor pattern offered by \pulselib{} is the \textit{muncher}.
Much like maximal munch algorithms~\cite{Cattell1980}, munchers are designed to reduce a graphical pulse representation to a linear data structure.
This is done through a combination of graph traversal and structural pattern matching.
A \pulselib{} muncher defines a template graph structure that is used to match against the input graph.
If the input graph matches the template, the muncher uses knowledge of this structure to extract key pulse parameters and format them into an alternative representation.
Otherwise, the muncher will raise an error indicating that the input graph is not compatible with the template.
In the context of diverse control systems, the role of munchers is two-fold.
First, they perform initial validation of the input graph.
If the provided graph does not match the expected structure, then it is not compatible with the target device, and raising an error is appropriate.
Second, munchers are used to format pulse parameters into a device-specific representation.
A device-aware muncher may be configured with device parameters (e.g., sampling rates, maximum frequency, etc.) that can be used to further validate and transform graph nodes into device-compatible data structures.

Finally, graph scheduling is a powerful feature of \pulselib{} that allows for representations of entire experiments as sequences of pulse graphs.
Multiple pulses on multiple channels can be choreographed through parallel and sequential scheduling contexts with unaddressed channels automatically padded with no-op \texttt{pulselib.Zero} nodes.
Coupled with unique channel identifiers, \pulselib{} schedules can model complex pulse experiments involving multiple distinct waveform generators.
Combining unique channel identifiers with device-aware munchers allows for the generation of device-specific pulse sequences from a single high-level description.

\subsection{Direct Digital Synthesizers}\label{sec:bg:dds}
As a demonstration of \pulselib{}'s utility, we design piplines for two devices common to quantum control systems: the Analog Devices AD9910~\cite{ad9910} controlled by ARTIQ~\cite{bourdeauducq_2016_51303} and the Xilinx Zynq UltraScale+ RFSoC~\cite{rfsoc} running Sandia's Octet firmware~\cite{Lobser2020}.
Both devices fall under the category of direct digital synthesizers (DDS).
Such devices are parameter-based signal generators that provide rapid and precise control over parameters of a sinusoid, and
they are used to generate microwave and radio frequency tones for coherent control of quantum systems~\cite{Clark2021,Kang2021}.

\begin{figure}
    \centering
    \includegraphics[width=\linewidth]{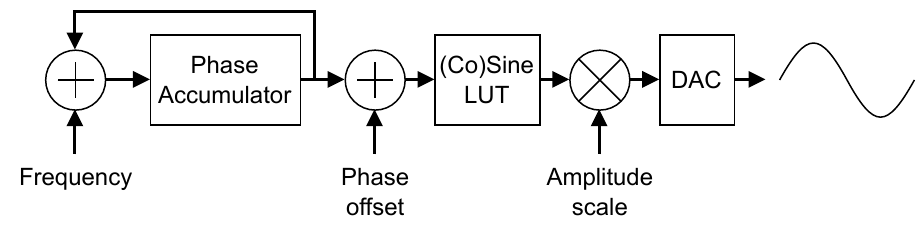}
    \caption{
        Block diagram of a DDS device.
        The core of a DDS consists of a phase accumulator, a sine look-up table, and a digital-to-analog converter (DAC).
        The phase accumulator generates a phase signal that is used to index into the sine (or cosine) look-up table (LUT) which returns the appropriate amplitude.
        After scaling, the DAC converts the digital output of the sine look-up table (LUT) into an analog signal.
    }
    \label{fig:dds-core}
\end{figure}

The common structure of a DDS is shown in \cref{fig:dds-core}.
A phase accumulator register is used to generate a phase signal that is used to index into a sine look-up table (LUT).
The rate of phase accumulation is controlled by a frequency register, which determines the phase increment each clock cycle.
After adding a phase offset, the accumulated phase value is mapped to amplitude via the sine LUT, which contains a digitized representation of a sine wave.
After amplitude scaling, its output is fed into a digital-to-analog converter (DAC) to produce the analog sinusoidal signal.
Hence, the DDS output is completely determined by the contents of the phase, frequency, and amplitude registers.
Register interactions via parallel or serial interfaces allow for rapid updates to the output signal, enabling fast frequency and phase hopping.

\subsubsection{ARTIQ AD9910}
The Advanced Real-Time Infrastructure for Quantum physics (ARTIQ) is a hardware-software ecosystem for real-time control of devices tailored for quantum information science~\cite{bourdeauducq_2016_51303}.
ARTIQ is designed to provide low-latency control of quantum devices, and it is capable of controlling a variety of hardware including arbitrary waveform generators (AWGs), field programmable gate arrays (FPGAs), and direct digital synthesizers (DDS).
Within the ARTIQ hardware ecosystem is the Urukul 9910, an expansion card sporting four channels, each equipped with an AD9910 DDS chip.
Each chip produces sinusoids of with frequencies up to 400 MHz with 14-bit resolution at 1 GS/s.
Frequency, phase, and amplitude control are provided through a set of registers programmable via SPI transactions, and discrete modulation envelopes can be obtained by programming the $1024\times 32$-bit internal RAM registers.

ARTIQ provides a thin abstraction layer over AD9910, exposing methods that initiate serial transactions for direct reading and writing of registers.
As a result, users must be aware of low-level details such as register widths, endianness, and timing constraints when programming the AD9910.
This leads to experiment code that is tightly coupled to the AD9910 hardware and difficult to reuse across different platforms.
Advanced AD9910 features such as discrete modulation further increase programming complexity since users must be aware of the internal RAM structure, how to program it, and how to initiate modulated pulse synthesis.

\subsubsection{Octet RFSoC}
The Xilinx Zynq UltraScale+ RFSoC is a system-on-chip (SoC) that integrates a field programmable gate array (FPGA) with multiple high-speed analog-to-digital and digital-to-analog converters (ADCs/DACs)~\cite{rfsoc}, and it has found extensive use in both trapped-ion~\cite{Clark2021,Sun2024} and superconducting~\cite{Stefanazzi2022a,Gartmann2024,Gebauer2023} quantum control systems.
For this work, we focus on the RFSoC's capabilities as a DDS enabled by the Octet firmware~\cite{Lobser2020}.
Octet leverages the RF capabilities of the RFSoC to realize a DDS-like device optimized for coherent control of trapped ions.
The Octet firmware configures the RFSoC with eight signal generating channels, each of which internally contains two DDS modules whose outputs are summed to produce the final signal.
For the remainder of this paper, ``tone'' refers to a single DDS module, while ``channel'' refers to the combined two-tone subsystem.

Like other DDS devices, Octet offers control over frequency, phase, and amplitude of each tone.
Internal spline engines enable discrete and smooth modulation of these parameters.
Where the Octet RFSoC surpasses traditional DDS devices is in phase control.
Each channel is equipped with two frame rotation registers that act as persistent phase accumulators.
Phase accumulated in these hardware registers may be applied to all subsequent frequency and phase changes, allowing for rapid frequency hopping while maintaining phase coherence.
This has applications in gate-level quantum computation where rotations about the $Z$-axis of the Bloch sphere amount to a simple phase shift.
Instead of driving such interactions with control fields, $Z$ rotations can be realized by a simple frame rotation.
This built-in phase tracking also has use in quantum simulation on trapped-ions where proper phase synchronization is needed when exciting many motional modes of ion chains~\cite{Sun2024}.
Additionally, this phase bookkeeping is done at the hardware level, freeing control software resources otherwise dedicated to phase tracking to other time-critical tasks such as responding to real-time feedback. 

A pulse-level quantum experiment on the Octet RFSoC is represented by a Jaqal~\cite{Morrison2020} file.
This file serves a role similar to header files in C programs; it specifies the sequences of gate and pulse operations to be executed on the RFSoC, but provides no implementation details on these pulses.
These details are instead provided by gate definition files written using JaqalPaw~\cite{Lobser2021}.
These files contain Python classes whose methods correspond to gate calls in a Jaqal files.
The return value of these methods are sequences of \texttt{PulseData}, a representation of an RFSoC channel pulse provided by JaqalPaw.
The Octet compiler then uses both the Jaqal file and gate definition file to generate a binary representation of the experiment that is uploaded to the RFSoC.

The workflow involving Jaqal files and gate definition files is appropriate for circuit-based quantum computation since experiment designers need only provide Jaqal file representations of their experiment while system maintainers are only concerned that gate definitions are optimal.
This separation of pulse definition and scheduling is not appropriate for pulse-level programming, where experiment designers require direct control of pulse parameters.
Further, parallel operations in Jaqal files are always synchronized, meaning that all pulses in a parallel operation start at the same time, limiting the expressibility of the Jaqal file representation.

\section{Transpilation Pipeline}\label{sec:pipeline}

\begin{figure}
    \centering
    \includegraphics[width=\linewidth]{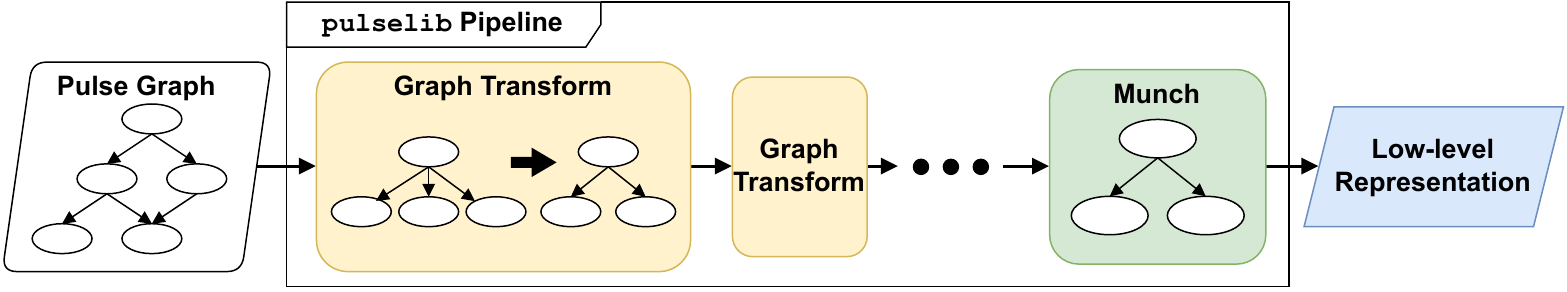}
    \caption{
        Pipeline for transpiling high-level pulse descriptions in \pulselib{} into low-level representations.
        The pipeline consists of a sequence of visitors and transformers that are applied to an input graph and ends with a muncher that outputs a low-level pulse representation if the processed graph matches the expected structure.
        }
    \label{fig:pulselib-pipeline}
\end{figure}

Graphs, schedules, and visitors together provide the framework for a transpilation pipeline through which high-level, system-agnostic pulse descriptions are transformed into low-level, device-specific representations, as shown \cref{fig:pulselib-pipeline}.
The input to this pipeline is the root node of a possibly parameterized graph representing an arbitrary pulse.
The pipeline consists of a sequence of visitors that are applied to an input graph and ends with a muncher that accepts the processed input graph and outputs a low-level pulse representation.

The key element of this pipeline is the muncher, as it is this component that asserts the constraints of the target device or representation.
In defining munchers, users specify sequences of functions that map acceptable input graphs to output data structures.
Concretely, these methods accept the root node of a graph and perform type checks on the root node and its children.
If these initial checks pass, then processing proceeds to the next layer of child nodes, and so on until it is confirmed that the entire graph meets the expected structure.
With full knowledge of the graph structure, munchers can then extract the necessary information from the graph and emit a low-level representation.
If any checks fail, the next function in the sequence is called, and so on until a match is found or all functions have been exhausted, at which point an error is raised.

Combining the above pipeline with the unique channel identifiers in \pulselib{} schedules allows for transpilation targeting complex control systems.
For example, each channel in a schedule may map to a unique pulse generating device, and a pipeline of transformations and munchers may be associated to each device.
When initiating schedule transpilation, channel information can be used to select the appropriate pipeline for each graph.
The output of this schedule-wide pipeline is then a mapping of channels to low-level representations at which point channel information can again be used to pass data to the next step in the target system's workflow (e.g., validation, compilation, upload, etc.).

\section{Implementation}\label{sec:imp}

\subsection{ARTIQ AD9910}\label{sec:imp:ad9910}
\begin{figure}
    \centering
    \includegraphics[width=.7\linewidth]{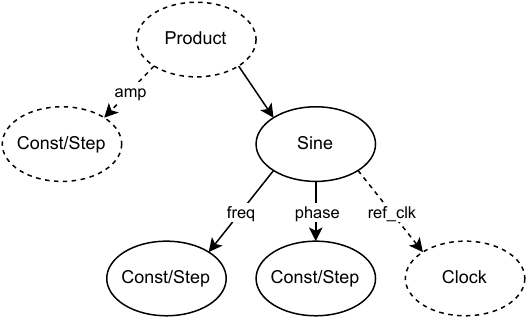}
    \caption{
        Template \pulselib{} graph capturing pulses synthesizable by an AD9910 DDS.
        Dashed nodes and edges are optional.
    }
    \label{fig:dds-template-graph}
\end{figure}
\Cref{fig:dds-template-graph} shows a template \pulselib{} graph capturing pulses synthesizable by an AD9910 DDS.
The core structure of the graph is built around a \texttt{pulselib.Sine} node which provides parameters for frequency, phase, and an optional reference clock.
Amplitude control is introduced by simply multiplying the \texttt{Sine} node with the desired amplitude, producing a \texttt{pulselib.Product} node.
Unity amplitude is inferred from the absence of a root \texttt{Product} node.
Pulse parameters are allowed to be simple constants or step functions, reflecting the AD9910's capability for discrete modulation of frequency, phase, and amplitude.
Step functions are represented as sequences of constants, each of which is represented by a \texttt{pulselib.Const} node.
The configurable phase behavior of the AD9910 is captured by the optional \texttt{pulselib.Clock} node.
Clock nodes are used internally by \pulselib{} for phase tracking, e.g., when converting to a sample-based representation~\cite{Dalvi2024}.
When present in a transpiled pulse graph, the presence of the clock enables phase continuous operation of the AD9910 in which accumulated phase is preserved between frequency changes.

With a template graph defined, the next step is to construct the visitors, transformers, and munchers that will extract the necessary information from the graph and convert it into a format suitable for the AD9910.
The transformation steps include removing all zero-duration nodes, replacing \texttt{pulselib.Cosine} nodes with \texttt{pulselib.Sine} nodes, and simplifying all arithmetic operator nodes with known operands.
Then, a visitor is applied which searches for the presence of a \texttt{pulselib.Clock} node and, if found, sets an internal flag to indicate phase continuous operation.

\begin{figure*}
    \centering
    \includegraphics[width=.8\linewidth]{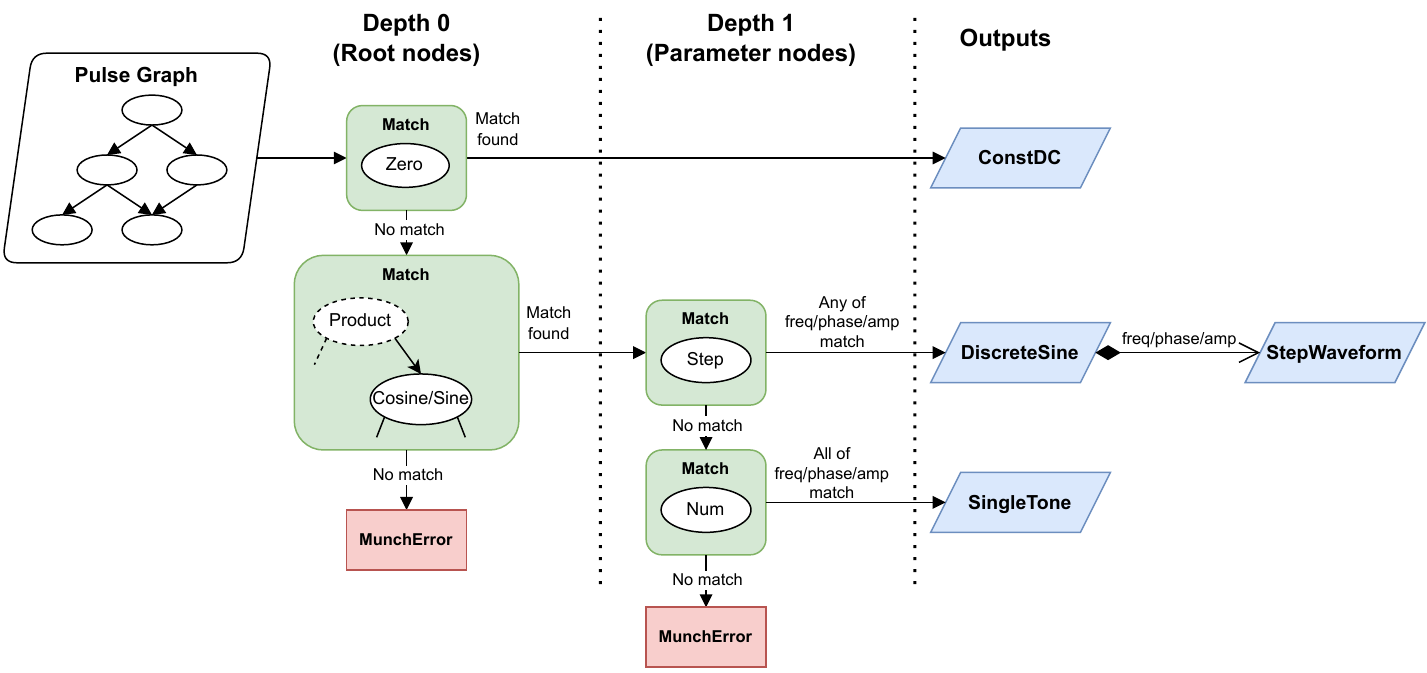}
    \caption{
        Muncher structure for the AD9910 IR.
        The muncher matches on two possible structures: a template graph and a no-op graph consisting of a single \texttt{pulselib.Zero} node.
        If a match with the template graph is found, a second layer of munchers is applied to leaf nodes to process frequency, phase, and amplitude parameters.
        This layer of munchers attempts to match step functions first and then constants.
        The data structure emitted when the template graph is successfully matched depends on the result of the second layer of munchers.
    }
    \label{fig:ad9910-muncher}
\end{figure*}

Finally, we enter the munching stage.
\Cref{fig:ad9910-muncher} shows the structure of the AD9910 muncher.
The implemented AD9910 muncher matches two possible structures: the template graph shown in \cref{fig:dds-template-graph} and no-op graph consisting of a single \texttt{pulselib.Zero} node.
The latter is required due to the automatic padding pulses inserted by \pulselib{} schedules.
We first attempt to match the \texttt{Zero} graph.
If this match succeeds, we immediately return a \texttt{ConstDC}, a data structure modelling a constant DC pulse, with zero amplitude.
If instead this match fails, we attempt to match the template graph.

Successful matching of the template graph triggers a second layer of munchers that are applied to leaf nodes of the template graph, i.e., the nodes corresponding to frequency, phase, and amplitude.
This layer of munchers attempts to match step functions first and then constants.
The step muncher matches on \texttt{pulselib.Sequence} nodes in which all children are \texttt{pulselib.Const} nodes and emits a \texttt{StepWaveform} data structure that is composed of a sequence of \texttt{ConstDC} objects.
If the step muncher fails, the constant muncher is called, which matches on \texttt{pulselib.Num} nodes and emits native Python \texttt{floats}.

The data structure emitted when the template graph is successfully matched depends on the result of the second layer of munchers.
If all three parameters are constants, the output is a \texttt{SingleTone} object which contains static frequency, phase, and amplitude values.
If any of the parameters are step functions, the output is instead a \texttt{DiscreteSine} object where the appropriate parameter is replaced with a \texttt{StepWaveform} object.

\subsection{Octet RFSoC}\label{sec:imp:rfsoc}
\begin{figure*}
    \centering
    \includegraphics[width=\linewidth]{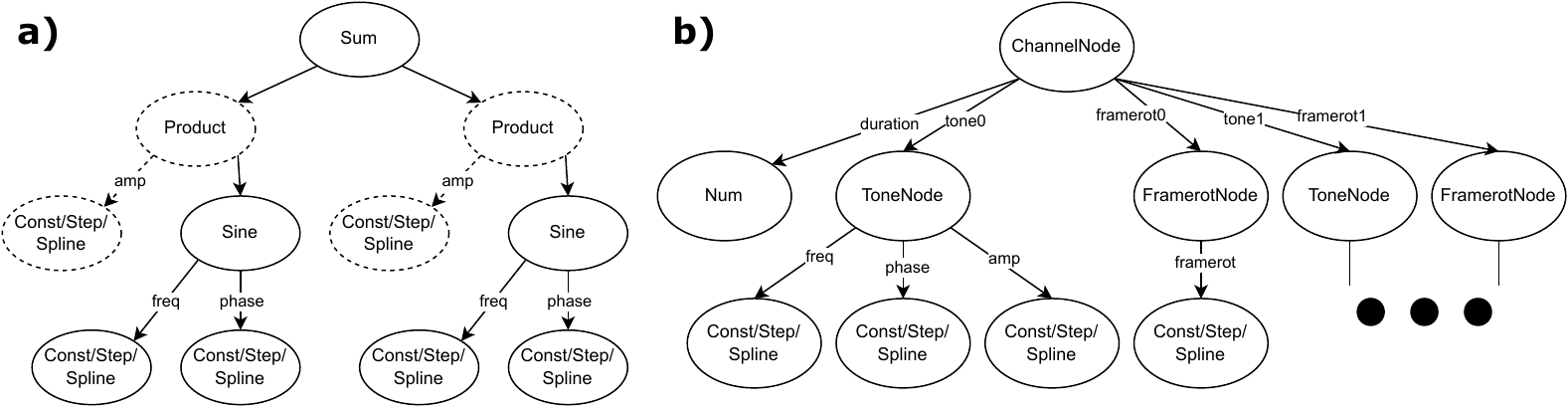}
    {\phantomsubcaption\label{fig:rfsoc-pl-graphs:base}}
    {\phantomsubcaption\label{fig:rfsoc-pl-graphs:optimized}}
    \caption{
        (a) Base \pulselib{} graph structure for RFSoC channels.
        (b) Optimized \pulselib{} graph structure for RFSoC channels.
        The \texttt{FramerotNode} and \texttt{ToneNode} nodes are used to model frame rotation operations and tone-specific information, respectively.
        The \texttt{ChannelNode} node combines two \texttt{FramerotNode} and two \texttt{ToneNode} nodes to form a complete RFSoC channel.
        Both classes contain metadata for advanced features such as phase synchronization, frame selection, and frequency feedback control.
    }

    \label{fig:rfsoc-pl-graphs}
\end{figure*}

Since the RFSoC feature set is a superset of the AD9910, the template graph of \cref{fig:dds-template-graph} can serve as a representation of an RFSoC tone.
The graphical representation of a full RFSoC channel is then two instances of an AD9910-like graphs connected by a \texttt{pulselib.Sum} node, as shown in \cref{fig:rfsoc-pl-graphs:base}.
Parameter nodes are allowed to represent constants, step functions, or spline curves, in keeping with the RFSoC's capabilities.

However, phase synchronization and frame rotations are advanced Octet features desirable in analog quantum simulation~\cite{Sun2024}.
Options such as phase synchronization may be inferred by the presence of a \texttt{pulselib.Clock} node as was done for the AD9910, but others such as frame rotations cannot.
We take advantage of \pulselib{}'s extensibility to introduce the family of custom nodes, shown in \ref{fig:rfsoc-pl-graphs:optimized}, that captures the full capabilities of the RFSoC.


We design the IR such that it is convertible to JaqalPaw's \texttt{PulseData} representation~\cite{Lobser2021} for compatibility with the Octet compiler.
A single \texttt{PulseData} object contains pulse data for both tones and both frames on a single channel.
Frame-specific data include frame rotation values and metadata flags for scheduling frame rotations and clearing accumulated phases.
Tone-specific data include non-duration pulse parameters with metadata flags for phase synchronization, frame selection, and frequency feedback control.
Pulse duration is a channel-wide parameter that applies to both tones and determines the step duration for discrete- or spline-modulated parameters.

\begin{figure*}
    \centering
    \includegraphics[width=.9\linewidth]{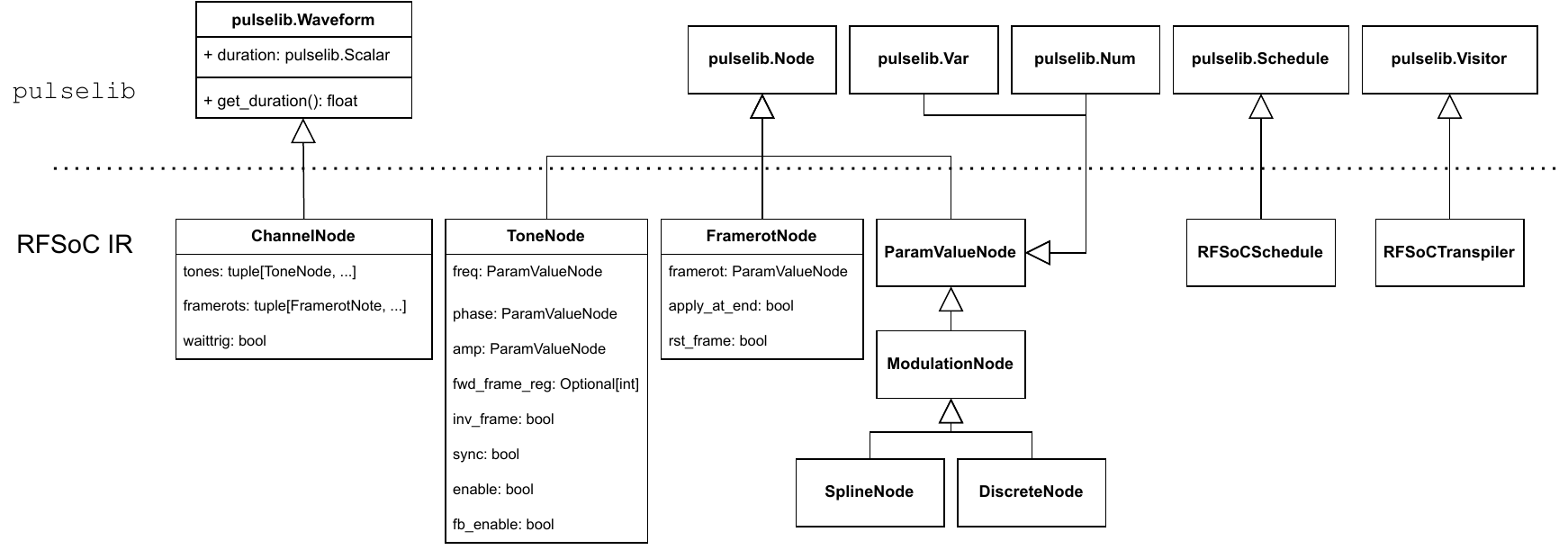}
    \caption{Inheritance structure of the RFSoC pulselib-based IR.}
    \label{fig:rfsoc-ir-inheritance}
\end{figure*}

\Cref{fig:rfsoc-ir-inheritance} shows elements of our RFSoC IR and their relationships to base \pulselib{} classes.
We first introduce \texttt{ParamValueNode}, an abstraction for values usable as pulse parameters.
Extending from \texttt{pulselib.Node} enables graphical representations of pulse parameters, and this is used by the \texttt{SplineNode} and \texttt{DiscreteNode} subclasses to model smooth and discretely modulated parameters respectively.
By registering them as virtual subclasses of \texttt{ParamValueNode}, both \texttt{pulselib.Num} (e.g., floating point literals) and \texttt{pulselib.Var} may be used as parameter values.

Instances of \texttt{ParamValueNode} may be used as parameters to \texttt{FramerotNode} and \texttt{ToneNode} objects.
\texttt{FramerotNode} models frame rotation operations and contains frame-specific metadata along with a \texttt{ParamValueNode} representing the frame rotation magnitude.
Similarly, \texttt{ToneNode} contains tone-specific metadata flags for phase synchronization, frame selection, and frequency feedback control in addition to \texttt{ParamValueNode} objects representing frequency, phase, and amplitude data.
Note that frame selection for a \texttt{ToneNode} is specification of a frame \textit{index} as opposed to the frame rotation operation itself.
This structure introduces a level of indirection that allows frame rotation operations to be re-used between channels.

Two instances of \texttt{ToneNode} and two of \texttt{FramerotNode} are combined in a \texttt{ChannelNode} to form a nearly complete channel pulse.
Total pulse duration and index of the targeted physical channel must be specified in order to fully convert \texttt{ChannelNode} objects to JaqalPaw \texttt{PulseData}.
Specification of pulse duration is deferred to the \texttt{ChannelNode} constructor since it is a channel-wide parameter.
Deriving \texttt{ChannelNode} from \texttt{pulselib.Waveform} simultaneously adds a duration parameter to its constructor, enables the graph representation, and allows parameterizable pulse durations.
To allow re-use of \texttt{ChannelNode} instances between physical RFSoC channels, specification of target channel index is inferred from the input schedule.

\begin{figure*}
    \centering
    \includegraphics[width=\linewidth]{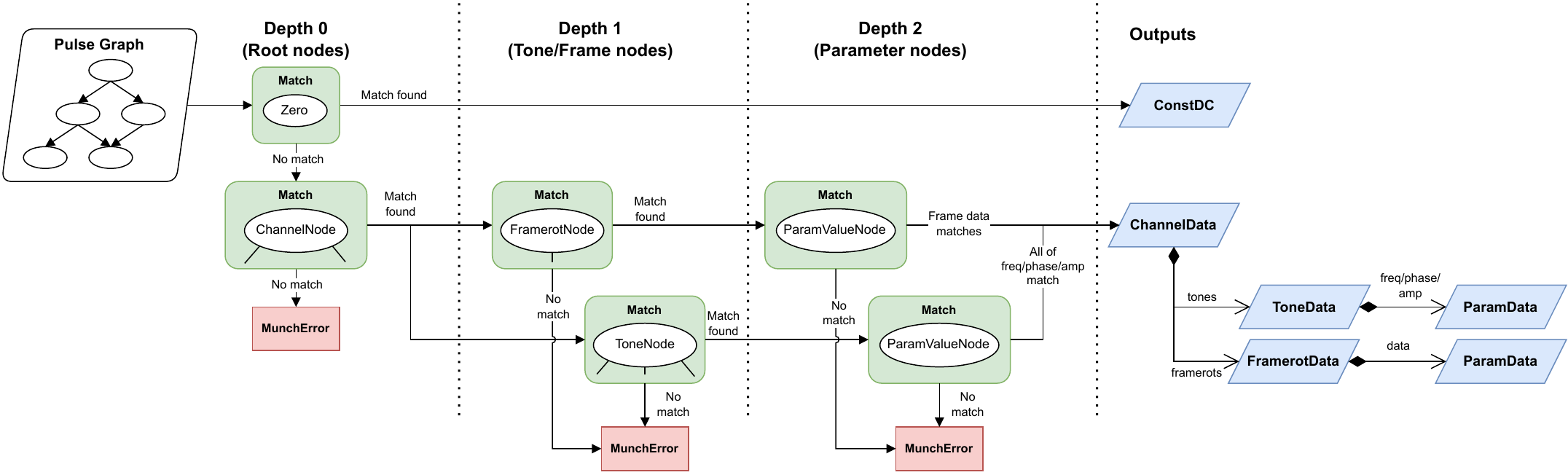}
    \caption{
        Muncher structure for the RFSoC IR.
        The first layer attempts to match either a no-op pulse represented as a \texttt{pulselib.Zero} node or a \texttt{ChannelNode}.
        A match of the former ends the munching process, producing a \texttt{ConstDC} object with zero amplitude.
        A match of the latter triggers a second layer of munchers that are applied to leaf nodes of the template graph, i.e., the nodes corresponding to frequency, phase, and amplitude.
        This second layer matches occurences of \texttt{FramerotNode} and \texttt{ToneNode} objects.
        This leads to the lowest layer in which tone and frame parameters modeled as \texttt{ParamValueNodes} are reduced to a data type compatible with JaqalPaw \texttt{PulseData}. 
        The final output in this case is a composite data structure \texttt{ChannelData} which internally contains \texttt{ToneData} and \texttt{FramerotData} objects.
        }
    \label{fig:rfsoc-ir-muncher}
\end{figure*}

\Cref{fig:rfsoc-ir-muncher} shows the structure of the RFSoC muncher.
In implementing the RFSoC muncher, we follow a similar approach to the AD9910 muncher.
Instead of converting the graph directly to \texttt{PulseData}, we first convert the graph to a \texttt{ChannelData} object, which is then converted to a \texttt{PulseData} object.
This approach allows us to reuse and quickly adapt the muncher should there be major modifications to the Octet compiler or \texttt{PulseData} representation.

The first layer of munchers attempt to first match the no-op pulse graph consisting of a single \texttt{pulselib.Zero} node and then the \texttt{ChannelNode} graph structure in \cref{fig:rfsoc-pl-graphs:optimized}.
If the latter match succeeds, the second layer checks for the presence of \texttt{FramerotNode} and \texttt{ToneNode} objects in the second layer of the graph.
If these checks pass, the muncher proceeds to the next layer of child nodes, which are expected to be \texttt{ParamValueNode} objects.
At this stage, the muncher attempts to convert \texttt{ParamValueNode} objects into a format recognized by \texttt{PulseData}.
Specifically, data from \texttt{SplineNodes} are formatted into tuples while data from \texttt{DiscreteNodes} are formatted into lists.
These outputs are inserted into the \texttt{ToneData} and \texttt{FrameData} data structures emitted by the second layer.
Finally, the \texttt{ToneData} and \texttt{FrameData} objects are combined into a \texttt{ChannelData} object, which is then emitted by the muncher.
Final conversion of \texttt{ChannelData} to \texttt{PulseData} requires a user-provided mapping of schedule channel identifiers to physical RFSoC channel indices.
Conversion then is as simple as mapping \texttt{ChannelData} attributes to \texttt{PulseData} constructor arguments.


As an aside, we note that the muncher structure shown in \cref{fig:rfsoc-ir-muncher} does not accept the base graph structure shown in \cref{fig:rfsoc-pl-graphs:base} as input.
Instead of adding munchers targeting this graph structure, we instead add transformation layers in front of the muncher that attempt to reduce the graph to the structure shown in \cref{fig:rfsoc-pl-graphs:optimized}, assuming default metadata parameters where needed.

\section{Benchmarks}\label{sec:bench}
We now turn to quantifying the performance of our \pulselib{}-based IRs.
We run benchmarks inspired by two common applications in trapped-ion quantum systems: sideband cooling (SBC)~\cite{Reed2024} and variational quantum algorithms (VQA)~\cite{Cerezo2021}.
We limit analysis to the RFSoC IR as two baseline IRs exist for comparison.
The first is PulseCompiler~\cite{Risinger}, an RFSoC programming IR built atop Qiskit Pulse~\cite{Alexander2020}, and the second is a custom representation that internally uses simple linear data structures instead of \pulselib{}'s graph-based representation.
This latter implementation, termed the ``direct approach'', sacrifices \pulselib{} features such as flexibility parametrizability in favor of simplicity and speed.
All benchmarks were run on an Intel i7-1255U CPU with 16 GB of RAM.

The direct approach implements a transpilation pipeline similar to \pulselib{} with the distinction that simple, linear data structures are used as internal representations.
These data structures form a hierarchy similar to that in the bottom half of \cref{fig:rfsoc-ir-inheritance} with the exception that variable pulse parameters are not supported.
A custom schedule implementation as also needed since \pulselib{} schedules cannot be used with objects outside \pulselib{}.
This direct schedule implementation copies the interface of \pulselib{} schedules, incorporating both sequential and parallel scheduling contexts with automatic zero padding.
Transpilation of these objects to JaqalPaw \texttt{PulseData} is performed by iterating over the schedule and mapping channel data abstractions directly to \texttt{PulseData}.

\begin{figure}
    \centering
    \includegraphics[width=.9\linewidth]{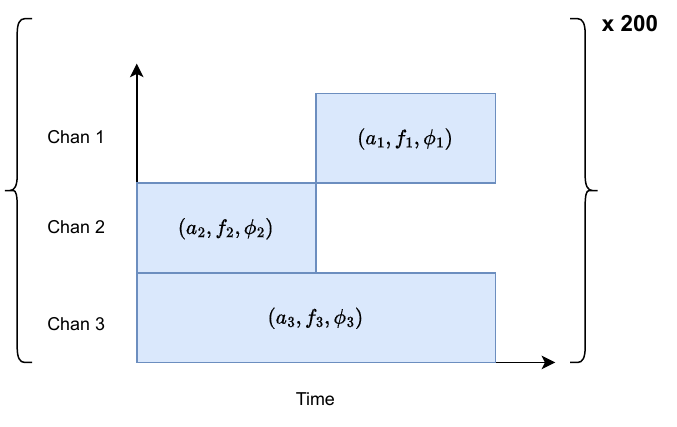}
    \caption{
        Structure of the SBC-inspired benchmark.
        Firts, a fixed unit schedule consisting of three pulses is constructed using one of three IRs.
        This unit schedule is then repeated 200 times to create the full schedule that is then transpiled to a lower-level representation compatible with the Octet compiler.
    }
    \label{fig:sbc-fig}
\end{figure}

\subsection{SBC-Inspired Benchmark}\label{sec:bench:sbc}
\Cref{fig:sbc-fig} shows the structure of the SBC-inspired benchmark.
We define a fixed unit schedule consisting of three non-parametrized pulses.
A single benchmark trial consists of
\begin{enumerate*}
    \item construction of pulse representations and the unit schedule,
    \item repetition of the unit schedule 200 times to create the full schedule, and
    \item transpilation of the full schedule to a lower-level representation compatible with the Octet compiler.
\end{enumerate*}

\begin{figure*}
    \centering
    \includegraphics[width=.9\linewidth]{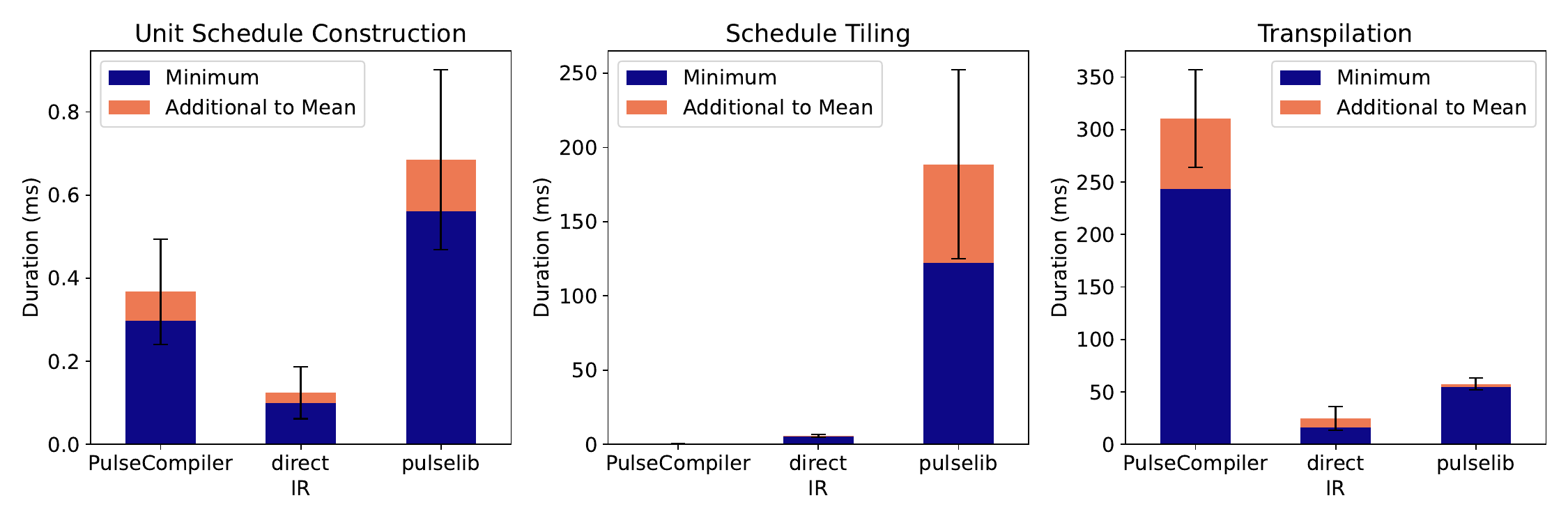}
    \caption{
        Minimum and mean execution times for (left) unit schedule construction (middle) unit schedule duplication and concatenation, and (right) transpilation of the full schedule for each IR.
        Error bars are standard deviation.
    }
    \label{fig:sbc-results}
\end{figure*}

Benchmark data collected over 500 trials for the \pulselib{}, direct, and PulseCompiler IRs are shown in \cref{fig:sbc-results}.
We first note PulseCompiler's efficient schedule representation but poor transpilation speed.
The former is especially apparent when tiling the unit schedule, where PulseCompiler requires at least 0.29 ms to construct the full schedule.
However, transpilation time requires at least 250 ms, an order of magnitude longer than the other IRs.
This contrasts \pulselib{}'s performance where full schedule construction requires at least 122 ms and transpilation at least 54 ms.
The large overhead in schedule construction for the \pulselib{} IR stems from graph post-processing.
After the schedule is finalized, equivalence transformations are performed on the graph to reduce it to a known form amenable to subsequent munching.
These transformations need only be applied once after the schedule is finalized, and since \pulselib{} schedules are immutable once built, this overhead is a one-time cost.
This leads to savings in transpilation, where \pulselib{} provides a best-case (minimum-over-minimum) speedup factor of 4.5 over PulseCompiler.
The direct approach, being a bespoke representation tightly coupled to the Octet RFSoC, shows the best performance with regard to schedule construction and transpilation, highlighting flexibility-performance trade-offs.

\subsection{VQA-Inspired Benchmark}\label{sec:bench:vqa}
\begin{figure}
    \centering
    \includegraphics[width=.9\linewidth]{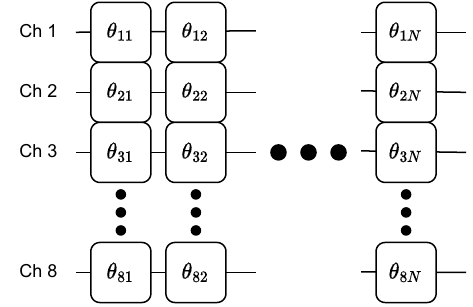}
    \caption{
        Parametrized structure of the VQA-inspired benchmark.
        The schedule consists of eight channels each of which containing $N$ parametrized pulses for a total of $8N$ parameters.
        A single trial of the benchmark consists of updating all $8N$ schedule parameters then transpiling the schedule to a lower-level representation compatible with the Octet compiler.
    }
    \label{fig:vqa-fig}
\end{figure}

The VQA-inspired benchmark is a more complex schedule that consists of a series of parametrized pulses.
\Cref{fig:vqa-fig} shows the structure of the VQA-inspired benchmark.
The schedule to transpile consists of $N$ layers of eight parametrized pulses each, leaving a total of $8N$ schedule parameters.
For \pulselib{} and the direct approach, pulse durations are parametrized while frequency is parametrized for PulseCompiler schedules due to a lack of support for parametrized pulse durations.
We stress here that the choice of pulse parameters is arbitrary since the substitution mechanism for each approach is independent of the pulse parameters, and the impactful metric here is instead the number of parameters in the schedule.
During each trial, all $8N$ schedule parameters are updated with randomly generated values and the updated schedule is transpiled.
Execution times are collected over 500 trials for each IR and schedule depth $N$.
Memory usage of the schedule at each schedule depth is calculated using the \texttt{pympler}~\cite{pympler} Python package which tracks the memory usage of complex Python objects.

\begin{figure*}
    \centering
    \includegraphics[width=\linewidth]{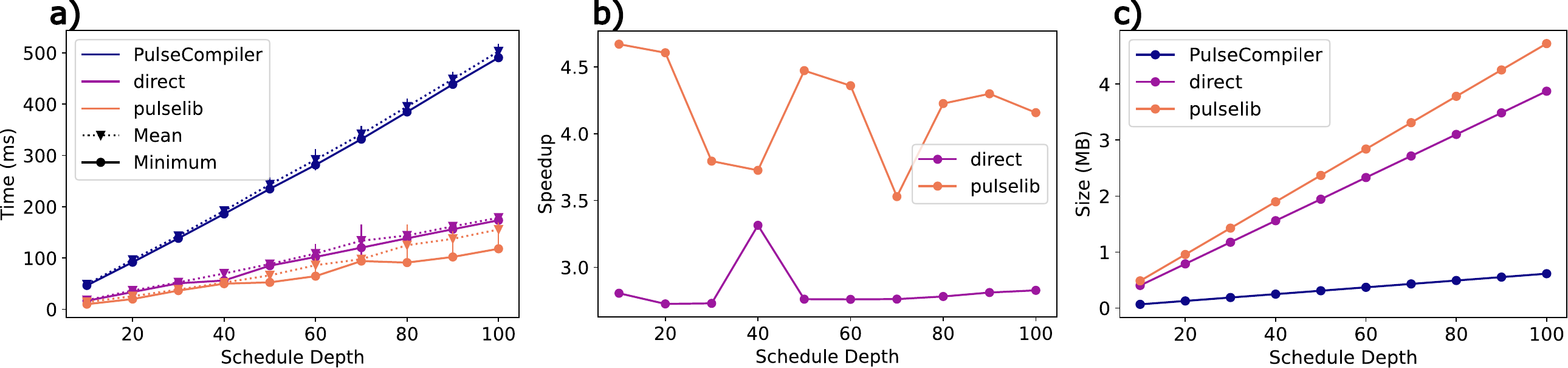}
    {\phantomsubcaption\label{fig:vqa-results:time}}
    {\phantomsubcaption\label{fig:vqa-results:speedup}}
    {\phantomsubcaption\label{fig:vqa-results:memory}}
    \caption{
        (a) Mean (dashed, triangles) and minimum (solid, circles) execution times for the VQA-inspired benchmark.
        The \pulselib{} implementation shows favorable scaling with larger parametrized circuits compared to PulseCompiler and even the direct approach.
        Error bars indicate standard deviation and may be smaller than the marker size.
        (b) Speedup of \pulselib{} and the direct approach over PulseCompiler using minimum execution times in (a).
        (c) Memory usage of full schedules for each IR.
        The \pulselib{} IR shows a moderate increase in memory usage per parameter compared to the direct approach, but this is offset by the increased performance.
        The PulseCompiler IR is the most memory efficient, but this comes at the cost of performance.
        Linear fits of memory usage curves reveal that \pulselib{} requires 5.90 kB per parameter, the direct approach 4.82 kB per parameter, and PulseCompiler 0.76 kB per parameter.
    }
    \label{fig:vqa-results}
\end{figure*}

\Cref{fig:vqa-results} gives the measured total execution time and consumed memory resources as the schedule depth $N$ is increased.
\Cref{fig:vqa-results:time} reveals that, for VQA-like schedules, \pulselib{} provides better best-case and average performance than the direct approach.
The improved performance of \pulselib{} relative to the SBC-like benchmark is due to \pulselib{}'s built-in support for parametrized schedules.
The \pulselib{} schedule makes heavy use of \texttt{pulselib.Var} nodes, which avoids the overhead of schedule reconstruction through schedule reuse between trials.
Instead, the schedule graph is traversed and the values of encountered \texttt{pulselib.Var} nodes are updated in-place.
To better compare the performance of \pulselib{} and the direct approach and compensate for systemic overhead (e.g. from background CPU processes), we calculate the best-case speedup (minimum-over-minimum) of the two relative to PulseCompiler in \cref{fig:vqa-results:speedup}.
This reveals that, for the same schedule depth, \pulselib{} can provide speedups up to 69\% larger than the direct approach.
This is accompanied by a moderate increase in memory usage, as shown in \cref{fig:vqa-results:memory}.
Using linear fits of the memory usage data, \pulselib{} requires 5.90 kB per parameter, an increase of 1.08 kB per parameter over the direct approach.

%
%



\section{Conclusions}\label{sec:conc}
In this paper, we have presented \pulselib{} as an ideal candidate for a system-agnostic pulse-level IR for programming quantum control systems equipped with a variety of signal generators.
As a demonstration, we have implemented IRs for the ARTIQ AD9910 and Octet RFSoC.
In the latter case, we took advantage of \pulselib{}'s extensibility to introduce custom nodes that capture the full capabilities of the RFSoC.
Using built-in \pulselib{} features like graph scheduling and munchers, we have shown how to construct a transpilation pipeline that can be used to convert high-level pulse representations into low-level device-specific representations and gave examples of how this can be done for the RFSoC and AD9910.
Finally, we have demonstrated the performance of our IRs through benchmarks inspired by common applications in trapped-ion quantum systems, showing that the time overhead of using \pulselib{} is comparable to or even less than that of existing IRs, particularly when using highly parametrized schedules.

A clear path forward is to expand the ecosystem of devices supported by \pulselib{}.
The current implementation is limited to the RFSoC and AD9910, but the extensibility of \pulselib{} allows for easy construction of pipelines targeting new devices.
For example, a \pulselib{} pipeline targeting  AWGs and other sample-based function generators would be considerably simpler than those targeting DDS devices since the former can synthesize arbitrary pulses.
As a result, the constraints on the structure of input pulse graphs can be relaxed, and the terminating muncher converts graphs to a discretized time-domain representation where parameters such as sampling rates, resolution, and interpolation techniques are determined by the targeted device.
Current \pulselib{} node primitives mirror common pulse shapes, e.g. \texttt{Sine}, \texttt{Gauss}, etc., offering a straightforward means of discretization.

While \pulselib{} is primarily intended as a description of pulses, its extensibility and graph-based representation allows for the introduction of new nodes that can be used to represent higher-level constructs.
For example, inspiration may be taken from abstract syntax tree (AST) implementations to incorporate control flow into \pulselib{}'s representation.
This would enable conditional pulse-level quantum experiments to be represented in \pulselib{}.
Given a control system that supports real-time feedback, e.g., ARTIQ~\cite{Riesebos2022}, this would allow for retargetable pulse-level experiments involving mid-circuit measurements, expanding \pulselib{}'s utility to quantum simulation~\cite{Koh2023} and error mitigation ~\cite{Botelho2022}.

Not considered in this work is the architecture for integrating \pulselib{} into a quantum control system.
For example, ARTIQ is a real-time control system, hence there is a degree of freedom in how transpiled pulse data is stored until it is needed.
In these real-time systems, a handle-based architecture~\cite{Riesebos2022,Alnas2024} may be used in which references to transpiled pulse data is given back to users after transpilation, and this handle is used to request synthesis from within real-time environments.
For systems that require compilation, such as the Octet RFSoC, \pulselib{} can serve as a substitute IR.
An entire experiment may be defined as a \pulselib{} schedule, and the outputs of the transpilation pipeline can be synthesized files (e.g., OpenQASM~\cite{Cross2022} or Jaqal~\cite{Morrison2020} files) compatible with the target compiler.

\section*{Acknowledgments}
The work was funded by the National Science Foundation (NSF) STAQ Project (PHY-2325080) and the NSF Quantum Leap Challenge Institute for Robust Quantum Simulation (OMA-2120757). Support is also acknowledged from the U.S. Department of Energy, Office of Science, National Quantum Information Science Research Centers, and Quantum Systems Accelerator.

\bibliographystyle{IEEEtran}
\bibliography{references}

\end{document}